\documentclass[conference]{IEEEtran}
\usepackage{graphicx,
	psfrag,
	epsfig,
	epstopdf,
    enumerate,
	amsthm,
    amsmath,
	amssymb,
	url,
	algorithm,
	algorithmic,
	balance,
	enumerate,
	color,
    url,
	tikz,
	pgfplots,
    geometry,
    placeins,
    cleveref,
    multirow,
    array,
}
\usepackage{textcomp}
\usepackage{amsmath}
\usepackage[nospace,noadjust]{cite}
\newgeometry{left=1.6cm,right=1.6cm,bottom=1.6cm, top=1.6cm}

\newcommand{\ignore}[1]{}

\begin{document}
\title{Optimized Random Forest Model for Botnet Detection Based on DNS Queries}
\author{
\IEEEauthorblockN{Abdallah Moubayed, MohammadNoor Injadat, and Abdallah Shami}
		
\IEEEauthorblockA{Electrical and Computer Engineering Department, Western University, London, Ontario, Canada \\
	emails: \{amoubaye, minjadat, abdallah.shami\}@uwo.ca
}
%
}
\maketitle

\begin{abstract}
The Domain Name System (DNS) protocol plays a major role in today's Internet as it translates between website names and corresponding IP addresses. However, due to the lack of processes for data integrity and origin authentication, the DNS protocol has several security vulnerabilities. This often leads to a variety of cyber-attacks, including botnet network attacks. One promising solution to detect DNS-based botnet attacks is adopting machine learning (ML) based solutions. To that end, this paper proposes a novel optimized ML-based framework to detect botnets based on their corresponding DNS queries. More specifically, the framework consists of using information gain as a feature selection method and genetic algorithm (GA) as a hyper-parameter optimization model to tune the parameters of a random forest (RF) classifier. The proposed framework is evaluated using a state-of-the-art TI-2016 DNS dataset. Experimental results show that the proposed optimized framework reduced the feature set size by up to 60\%. Moreover, it achieved a high detection accuracy, precision, recall, and F-score compared to the default classifier. This highlights the effectiveness and robustness of the proposed framework in detecting botnet attacks.
\end{abstract}
\section{Introduction}\label{Intro_dns}
\indent The Domain Name System (DNS) protocol plays a major role in today's Internet as it translates between website names and corresponding IP addresses \cite{DNS_definition}. Additionally, it is also used to locate servers and mailing hosts which directly impacts the data exchange across the Internet \cite{DNS_definition}.\\
\indent However, due to the lack of processes for data integrity and origin authentication, the DNS protocol is prone to various security threats and potential attacks \cite{DNS_vulnerabilities,DNS_vulnerabilities1}. One recent example is the  distributed denial of service (DDoS) attack on Dyn in October 2016 \cite{DNS_attacks1,DNS_attacks2}. As a result of this attack, a large portion of Internet services in America went down \cite{DNS_attacks1,DNS_attacks2}. One class of attacks that relies on the naivety of the DNS protocol is the botnet attack class. This class of attacks is particularly dangerous as it can lead to various undesired consequences such as network outages, user information leakage, and data privacy issues \cite{botnet_danger,botnet_attack}. Accordingly, researchers have proposed multiple different mechanisms to mitigate and eliminate these attacks \cite{SDP1,SDP2}. One such mechanism is the DNS over HTTPS (DoH) protocol. This protocol encrypts DNS queries and sends them in cover tunnels to enhance privacy and combat man-in-the-middle attacks \cite{DoH1}. Despite its promise, this protocol still cannot fully protect DNS queries due to attackers initiating malicious tunnels themselves. Therefore, further effective and efficient detection mechanisms are needed on top to ensure that systems are properly protected.\\
\indent One promising solution to detect DNS-based botnet attacks is adopting machine learning (ML) based solutions. ML allows systems to “learn” without being told what to do, making them dynamic and able to adapt new inputs \cite{Moubayed_EDM}. Furthermore, given that ML classification techniques have proved to be effective and efficient in a variety of applications including network security and intrusion detection frameworks \cite{Moubayed_EDM1,Moubayed_EDM2,Injadat_EDM1,Injadat_EDM2,Injadat_BO,Injadat_IDS1,Injadat_IDS2}, they are a prime candidate to be deployed for effective detection of botnets. Additionally, it is crucial to optimize these ML-based detection models rather than use default versions to ensure that the solution is performing to its maximum capacity \cite{Li_HPO,Moubayed_thesis,Injadat_thesis}.\\
\indent To that end, this paper proposes the use of optimized ML models to detect botnet attacks based on their DNS queries. More specifically, the proposed framework consists of using information gain as a feature selection method to reduce the computational complexity of the resulting model (by reducing the feature space) and genetic algorithm (GA) as a hyper-parameter optimization model to tune the parameters of a random forest (RF) classifier. Note that the choice of methods was motivated by the promising performance they have exhibited in previous ML-based network security frameworks.

\indent The main contributions of this work are:
\begin{itemize}
	\item \textit{Proposing} a novel optimized ML-based framework for DNS-based botnet detection with reduced computational complexity and enhanced detection performance.
	\item \textit{Evaluating} the performance of the proposed model using state-of-the-art TI-2016 DNS dataset. 
\end{itemize}
To the best of our knowledge, no previous work focusing on DNS-based botnet detection using the aforementioned dataset has proposed such an optimized and computationally efficient framework.\\
\indent The remainder of this paper is organized as follows: Section \ref{related_work_dns} provides a brief overview of the previous works from the literature. Section \ref{proposed_approach_dns} presents the proposed Optimized botnet detection framework and determines its complexity. Section \ref{dataset_description_dns} provides a description of the dataset considered. Section \ref{results_dns} describes the experiment setup and discusses the obtained results. Finally, Section \ref{conc_dns} concludes the paper.
\section{Related Work}\label{related_work_dns}
\indent As mentioned earlier, ML has been heavily proposed as a promising and effective solution in network security and thus has garnered significant attention \cite{Injadat_BO}\cite{Injadat_IDS1}\cite{Injadat_IDS2}\cite{Li_IDS}. For example, Injadat \textit{et al.} proposed the use of different feature selection and hyper-parameter optimization models to reduce the complexity and optimize the parameters of different ML classifiers for network intrusion detection with their experimental results showing that the proposed models achieved high detection accuracy and low false alarm rate \cite{Injadat_BO,Injadat_IDS1}. Similarly, Salo \textit{et al.} proposed a clustering-enabled classification model using ensemble feature selection for intrusion detection \cite{Injadat_IDS2}. Experimental results showed that the proposed framework was successful in accurately detecting unseen attack patterns \cite{Injadat_IDS2}. On the other hand, Li \textit{et al.} proposed the use tree-based ML models to detect both intra-vehicle and inter-vehicle intrusions in autonomous and connected vehicles with their proposed model having high detection accuracy and a low computational complexity \cite{Li_IDS}. \\
\indent Within the context of DNS, multiple researchers have proposed the use of ML models for malicious DNS queries detection \cite{Moubayed_DNS1,Moubayed_DNS2,R2_1,s64a,s66,s67}. For example, Moubayed \textit{et al.} proposed the use of exploratory data analytics, ensemble feature selection models, and ensemble classification models to understand the characteristics of different DNS typo-squatting features and accurately detect malicious URLs respectively \cite{Moubayed_DNS1,Moubayed_DNS2}. On the other hand, Sivakorn \textit{et al.} proposed using deep neural network models to detect malicious DNS queries \cite{R2_1}.\\
\indent In the context of botnet detection, McDermott \textit{et al.} proposed the use of a Bidirectional Long Short Term Memory based Recurrent Neural Network (BLSTM-RNN) to detect botnet activity within IoT devices and networks \cite{s64a}. Experimental results demonstrated that the proposed model achieved a detection accuracy ranging between 92\%-99\%. On the other hand, Pektas \textit{et al.} considered three different ML algorithms, namely RF, logistic regression (LR), and support vector machines (SVM) to detect botnet attacks during network flow analysis \cite{s66}. Experimental results illustrated that the proposed model achieved high classification accuracy for both botnet and normal traffic instances. Similarly, Chen \textit{et al.} also investigated the used of ML models to detect botnets \cite{s67}. The authors proposed a conversation-based network traffic analysis method to identify malicious botnet traffic. Experimental results showed that the proposed model resulted in a 13.2\% decrease in the false positive rate of botnet traffic detection. Furthermore, it was shown that the RF algorithm had a high detection accuracy (93.6\%) and a low false positive rate (0.3\%).\\ 
\indent These works further illustrate the huge potential and effectiveness of ML-based solutions for botnet detection. Nonetheless, most of the works focusing on ML-based malicious activity detection for DNS protocol use default parameters for the classification models. Thus, it is worth exploring optimized ML models to improve the effectiveness of such frameworks.
\section{Proposed Approach}\label{proposed_approach_dns}
\subsection{Proposed Approach Description}
\indent This paper proposes the use of information gain feature selection, GA hyper-parameter optimization, and RF classification model as part of the optimized ML-based framework for botnet detection. The proposed approach, as shown in Fig. \ref{dns_approach_fig}, can be divided into three components:
\begin{figure}[!t]
	\centering
	\includegraphics[scale=0.35]{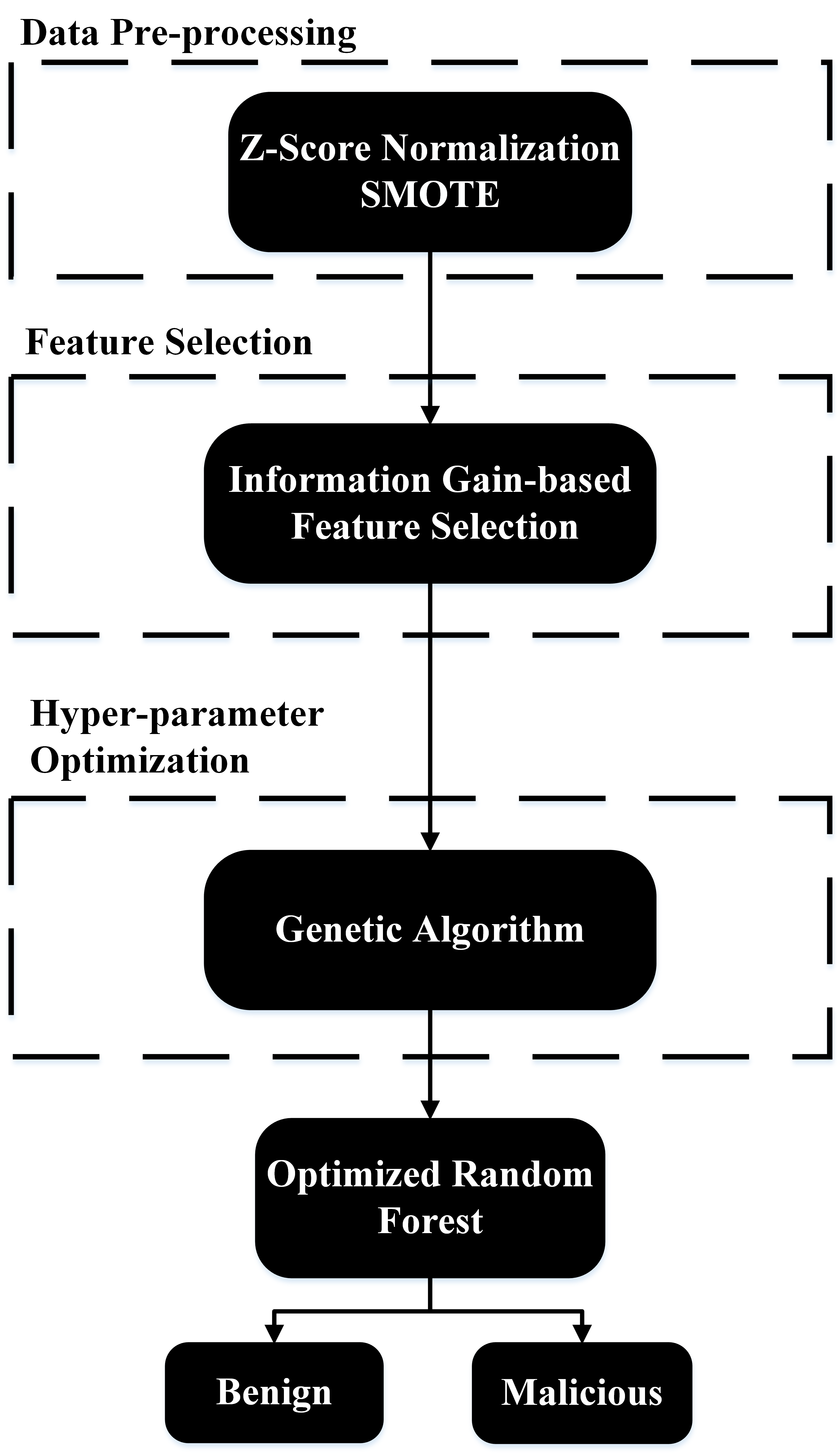}
	\caption{Proposed Optimized RF-based Botnet Detection Framework}
	\label{dns_approach_fig}
\end{figure}
\begin{enumerate}
	\item Data pre-processing: This stage is composed of two steps. The first is applying feature normalization using Z-score method. This is done so that all features considered have a similar dynamic range rather than one feature dominating due to its large dynamic range \cite{Injadat_IDS1}. The second step is applying Synthetic Minority Oversampling TEchnique (SMOTE) to tackle the class imbalance problem often present in network security datasets by oversampling the minority class \cite{imbalance1}. This can help improve the performance of the classification model and reduce the training sample size due to having enough samples to understand the behavior of each class \cite{SMOTE1}. 
	\item Feature selection: The goal at this stage is to reduce the number of features inputted to the ML model to help reduce its computational complexity while still maintaining or even improving its detection performance \cite{FS1}. To achieve this, information gain method is used to select the relevant features by ranking them according to the amount of information (in bits) they provide about the class \cite{IGBFS}.  
	\item Hyper-parameter Optimization: This stage focuses on optimizing the hyper-parameters of the ML classification model to improve its performance.  As such, this work proposes the use of GA method to optimize the hyper-parameters of the RF model. GA method is a well-known meta-heuristic algorithm used to determine high quality solutions to combinatorial optimization problems using biologically-inspired operations including mutation, crossover, and selection \cite{GA1}. These operations enable the efficient search for a suitable solution within the search space \cite{GA1}. Therefore, the GA method is proposed due to its near-optimal performance and low complexity \cite{GA1}. 
\end{enumerate}
\subsection{Complexity of Proposed Approach}
\indent The complexity of the proposed framework is dictated by the complexity of each stage. Assume that the dataset consists of $M$ samples and $N$ features. The complexity of the data pre-processing stage is governed by the complexity of the Z-score normalization method and the SMOTE method. The Z-score method has complexity of $O(N)$ given that it normalizes all the features using the respective means and standard deviation. In contrast, the SMOTE method has a complexity of $O(M^2_{min}N)$ where $M_{min}$ is the number of minority class instances \cite{SMOTE_complexity}. The complexity of the feature selection stage is that of the information gain method. This method has a complexity of $O(MN)$ since this method calculates the class-feature joint probabilities to determine the relevant features \cite{IGBFS_complexity}. Thirdly, the complexity of the GA method is $O(N_{pop}N_{parm})$ where $N_{pop}$ is the population size (\textit{i.e.} number of chromosomes initially assumed) and $N_{parm}$ is the number of hyper-parameters to be optimized \cite{GA_complexity}. Finally, the complexity of the optimized RF classification model is $O(M^2\sqrt{N_red}t)$ where $t$ is the number of trees within the RF classifier. However, the training time can be significantly reduced to approximately $O(\frac{M^2\sqrt{N_red}t}{threads})$ where $threads$ is the maximum number of participating threads since this classifier allows for multi-threading \cite{Li_IDS}.
Therefore, the complexity of the proposed framework is $O(\frac{M^2\sqrt{N_red}t}{threads})$. This highlights the computational efficiency of the proposed framework as it uses a reduced feature set and multiple threads.
\section{Dataset Description}\label{dataset_description_dns}
\begin{figure}[!h]
	\centering
	\includegraphics[scale=.5]{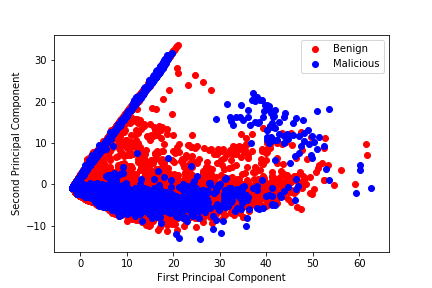}
	\caption{First and Second Principal Components of TI-2016 DNS Dataset}
	\label{pca}
\end{figure}

\indent The dataset considered in this work is the TI-2016 DNS dataset \cite{DNS_botnet_dataset}. To generate the dataset, the authors collected a real-world DNS traffic from more than 4000 active users during peak hours across 10 random days between April-May 2016. From the collected network traffic, a series of 25 features were extracted including request-based, domain-based, response-based, IP-based, and mapping features in addition to 1 class feature.  Accordingly, the resulting dataset consists of 601,092 \textbf{benign} and 7,644 \textbf{malicious} instances, meaning that it is significantly imbalanced. Fig. \ref{pca} plots the first and second principal components of the dataset, illustrating its non-linear nature and further emphasizing the class imbalance problem. 
\section{Experiment Results \& Discussion}\label{results_dns}
\indent The experiment results are divided into two parts, namely the feature selection and the optimized classification model results.
\subsubsection{{Feature Selection}}
\begin{figure}[!t]
	\centering
	\includegraphics[scale=.8,angle =-90]{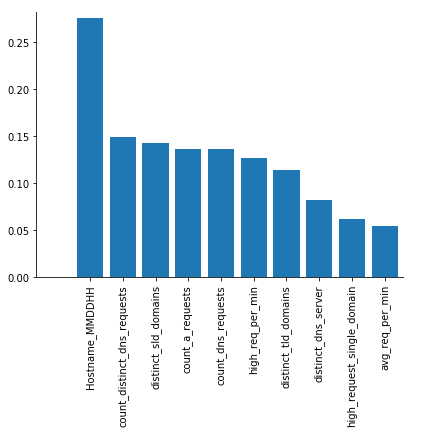}
	\caption{Mutual Information Score of Top 10 Features of TI-2016 DNS Dataset}
	\label{mi_fig}
\end{figure}

\indent Starting with the feature selection results, Fig. \ref{mi_fig} shows the mutual information score of the different features. Based on these score, a group of 10 features are chosen, namely: Host name and timestamp, number of DNS requests, number of distinct DNS requests,
highest number of requests (single domain), average number of requests per minute, highest number of requests per minute, number of Type A DNS requests, number of distinct top level domains queried, number of distinct second level domains queried, and number of distinct DNS servers queried. This represents a 60\% reduction in the feature set size. 
\subsubsection{{Optimized RF Classification Model Performance}}
\indent To evaluate the impact of the proposed optimized RF-based framework for botnet detection, its performance is compared to a default RF classifier that was applied to the original complete dataset without any feature selection or instance oversampling. As shown in Table \ref{labeled_dataset_results}, the proposed optimized RF-based framework outperforms the default RF classifier. Although the default RF classifier achieved a higher accuracy of 98.85\%, it had lower precision, recall, and F-score values. In particular, the default RF classifier had a precision, recall, and F-score of 0.92, 0.55, and 0.59 respectively for the malicious target class. This means that the default RF classifier is missing many of the actual malicious hosts, thereby being less effective for the problem at hand. This is attributed to the imbalanced nature of the original dataset under consideration which does not favor the development of a robust learner. In contrast, the proposed optimized RF-based framework still maintained a high botnet detection accuracy of 94.71\%. Moreover, it achieved a value of 0.95 for precision, recall, and F-score. The high precision and recall values show that the proposed optimized RF-based framework is robust and effective in detecting malicious hosts by being both sensitive and specific to malicious behavior. This emphasizes the effectiveness of the proposed framework in detecting botnet attacks from malicious hosts.  
\begin{table}[!t]
	\centering
	\caption{Performance Evaluation of Classifiers}
	\scalebox{0.85}{
		\begin{tabular}{|p{2.7cm}|p{1.4cm}|p{1.4cm}|p{1.4cm}|p{1.1cm}|}	\hline
			Algorithm & Accuracy (\%) & Precision & Recall&F-score\\ \hline
			Default RF&\textbf{98.85}&0.92&0.55&0.59\\ \hline
			Optimized RF-based Framework&94.71&\textbf{0.95}&\textbf{0.95}&\textbf{0.95}\\ \hline
		\end{tabular}
	}
	\label{labeled_dataset_results}
\end{table}
\section{Conclusion \& Future Works}\label{conc_dns}
\indent The Domain Name System (DNS) protocol plays a major role in today's Internet as it translates between website names and corresponding IP addresses. However, it is prone to various security threats and potential attacks due to the lack of data integrity and origin authentication processes. One class of attacks that relies on the naivety of the DNS protocol is the botnet attack class. Researchers have proposed different mechanisms to mitigate and eliminate these attacks. Despite the promising mechanisms previously proposed, further effective and efficient detection mechanisms are needed on top of existing ones to ensure that systems are properly protected.\\
\indent To that end, this paper proposed an optimized RF-based framework to detect botnet attacks based on their DNS queries. More specifically, the proposed framework used information gain as a feature selection method and genetic algorithm (GA) as a hyper-parameter optimization model to tune the parameters of the RF classifier. The proposed optimized RF-based framework reduced the feature set size by up to 60\%. Moreover, it achieved a high detection accuracy, precision, recall, and F-score values of 94.71\%, 0.95, 0.95, and 0.95 respectively. This emphasized the effectiveness of the proposed framework in detecting botnet attacks from malicious hosts.\\
\indent To extend this work, multiple research directions exist. One direction is extending the framework to consider multiple attacks rather than merely classifying the data in a binary manner. Another potential direction worth exploring is investigating hybrid ML models for online botnet detection frameworks. A third potential research direction is considering non-numerical features as part of any botnet detection models since such features may contain valuable information.

\small
\bibliographystyle{IEEEtran}
\bibliography{Ref1}
\end{document}